\documentclass[twocolumn]{svjour3}
\usepackage{graphicx}
\usepackage{epstopdf}
\usepackage{amsmath}
\usepackage{epsfig}
\usepackage{tensor}
\usepackage{amsfonts}
\usepackage{subfigure}
\usepackage{xcolor}
\usepackage{setspace}
\usepackage{listings}
\usepackage{soul}

\newcommand{\pd}[2]{\ensuremath{\frac{\partial #1}{\partial #2}}}

\newcommand{\epdt}[1]{\ensuremath{\dot{\epsilon}_i^{#1}}}
\newcommand{\notethis}[1]{#1}

\numberwithin{equation}{section}

\title{Kinematic flow patterns in slow deformation of a dense granular material}
\author{Koushik Viswanathan \and Anirban Mahato \and Tejas G. Murthy \and Tomasz Koziara \and Srinivasan Chandrasekar}
\institute{K. Viswanathan \at Center for Materials Processing and Tribology\\Purdue University, West Lafayette, IN, USA\\\email{kviswana@purdue.edu} 
\and
A. Mahato \at Center for Materials Processing and Tribology\\Purdue University, West Lafayette, IN, USA\\
\and
T. G. Murthy \at Dept. of Civil Engineering\\ Indian Institute of Science, Bangalore, India\\
\and
T. Koziara \at School of Engg. and Computing Sciences\\ Durham University, Durham, UK\\
\and
S. Chandrasekar \at Center for Materials Processing and Tribology\\Purdue University, West Lafayette, IN, USA}

\begin{document}
\maketitle

\begin{abstract}
The kinematic flow pattern in slow deformation of a model dense granular medium is studied at high resolution using \emph{in situ} imaging, coupled with particle tracking. The deformation configuration is indentation by a flat punch under macroscopic plane-strain conditions. Using a general analysis method, velocity gradients and deformation fields are obtained from the disordered grain arrangement, enabling flow characteristics to be quantified. The key observations are the formation of a stagnation zone, as in dilute granular flow past obstacles; occurrence of vortices in the flow immediately underneath the punch; and formation of distinct shear bands adjoining the stagnation zone. The transient and steady state stagnation zone geometry, as well as the strength of the vortices and strain rates in the shear bands, are obtained from the experimental data. All of these results are well-reproduced in exact-scale Non-Smooth Contact Dynamics (NSCD) simulations. Full 3D numerical particle positions from the simulations allow extraction of flow features that are extremely difficult to obtain from experiments. Three examples of these, namely material free surface evolution, deformation of a grain column below the punch and resolution of velocities inside the primary shear band, are highlighted. The variety of flow features observed in this model problem also illustrates the difficulty involved in formulating a complete micromechanical analytical description of the deformation.
\PACS{83.80.Fg, 81.20.Ev, 81.05.Rm, 81.40.Lm}
\keywords{dense flow, vortices, stagnation, shear bands, pattern formation}
\end{abstract}


\section{Introduction}
\label{sec:introduction}

Flow patterns observed in granular materials can resemble those of fluids as well as solids, depending on the density \cite{SarkarETAL_PhysRevLett_2013}. At low densities, the flow is well described by kinetic theory \cite{BrilliantovPoschel_KineticTheoryGranular,AransonTsimring_GranularPatterns}, corresponding to the hydrodynamic regime. As the density is increased, long--lasting interparticle contacts occur, resulting in solid--like behavior, characterized by force chains \cite{ZhangETAL_GranularMatter_2010}. This is the case, for instance, in the deformation of a dense granular pile under the action of gravity, where the contacts between adjacent particles last for infinite time. 

When the particles constituting the medium are extremely small, the deformation can be described by a suitable average over a Representative Volume Element (RVE). In this framework, the material is said to yield locally, whenever the shear stress is high \cite{Nedderman_StaticsKinematicsGranular}. However, as the particle size increases, the corresponding size of the necessary RVE also increases, and such continuum methods become inaccurate \cite{RycroftETAL_JMechPhysSolids_2009}. At the level of an individual particle or small clusters of particles, plastic flow initiation is linked to the density of collective structures. Examples of these are Shear Transformation Zones (STZs) \cite{Argon_ActaMet_1979,FalkLanger_PhysRevE_1998}, localized fluidization spots \cite{KamrinBazant_PhysRevE_2007} and deformation/ crushing of the individual particles \cite{Nedderman_StaticsKinematicsGranular}. 

Apart from obvious interest to problems in soil mechanics, the flow of dense granular materials is central to tablet processing in the pharmaceutical sector \cite{Duran_SandsPowdersGrains} and powder deformation processing of metals and ceramics \cite{Fleck_JMechPhysSolids_1995}. Dense granular materials have also served as a model for studying shear banding instabilities \cite{Spaepen_ActaMet_1977,Argon_ActaMet_1979,FalkLanger_PhysRevE_1998,JiangDai_JMechPhysSol_2009,ShimizuETAL_MatTrans_2007,BiETAL_Nature_2011} in amorphous metals (Bulk Metallic Glasses or BMGs). In all of these applications, the key problem is that of predicting deformation and flow patterns under a specified kinematic configuration. Besides enabling improved understanding of energy dissipation and losses during flow, such quantitative information will also help in developing methods to control the level of homogeneity of the resulting material microstructure.

The focus of the present study is on characterizing dense flow in the mesoscopic regime, where the constituent particles are large --- about $1/25^{th}$ a typical macroscopic problem length --- in contrast to a material such as sand, where the flow can be well approximated as a continuum. For this purpose, we use the model experimental system of punch indentation, with corresponding $1:1$ numerical simulations. The selection of this system is motivated by the varied flow features observed in punch indentation with metals as well as sand \cite{Nedderman_StaticsKinematicsGranular,Hill_MathematicalTheoryPlasticity,TordesillasShi_ProcRoySocA_1999,PetersETAL_PhysRevE_2005,KondicETAL_PhysRevE_2012,MurthyETAL_ProcRoySocA_2014}. The system captures many features common to deformation and powder processing. A flat--ended rectangular punch is used to penetrate the surface of the material, under macroscopic plane--strain conditions. Microscopically, however, each grain can move in all three dimensions. The flow is characterized at high--resolution using \emph{in situ} imaging and particle tracking techniques. In order to concentrate solely on the role of structural rearrangements, the effects of particle deformation and polydispersity are suppressed by the use of nearly rigid monodisperse spherical grains. Our results indicate that the observed flow patterns are qualitatively unchanged even if these constraints are relaxed.

Concurrent with the experiments, the use of numerical simulations allows us to probe kinematic details that are not easily accessed in the imaging setup. To this end, we use the Non Smooth Contact Dynamics (NSCD) method \cite{Jean_CompMethApplMechEngg_1999,Radjai_MechMat_2009,PoschelSchwager_CompGranDynamics} to handle the long lasting interparticle contacts that occur in dense flow. Mimicking the experiments, simulations are performed with monodisperse rigid spherical particles. The laws governing individual particle dynamics are simple and the observed flow patterns result solely from collective motion of the grains. \notethis{Due to fundamental differences between granular packings in two and three dimensions \cite{WeaireAste_PursuitPerfectPacking}, 3D simulations are necessary to handle monodispersity without causing crystallization.}

With this as background, we describe the experimental setup and numerical simulation method in Sec.~\ref{sec:methods}. Our main results are presented in Sec.~\ref{sec:results}. The primary flow features are then described and analyzed; these are quantitatively reproduced by the numerical simulations. \notethis{In Sec.~\ref{sec:discussion}, we discuss the origin and characteristics of the flow patterns and draw comparisons with similar features seen in other experiments, notably in punch indentation of metals}. The results and their implications are summarized in Sec.~\ref{sec:conclusions}.


\section{Methods}
\label{sec:methods}

In order to quantify effects of structural rearrangements on the kinematic flow pattern in a dense non--cohesive granular medium, model experiments  were carried out using plane--strain punch indentation. Corresponding numerical simulations were performed using the Non--Smooth Contact Dynamics (NSCD) method. 

\subsection{Experimental setup}

\begin{figure}
\clearpage
  \centering
  \mbox
  {
    \subfigure[\label{fig:schematic}]{\includegraphics[scale=0.45]{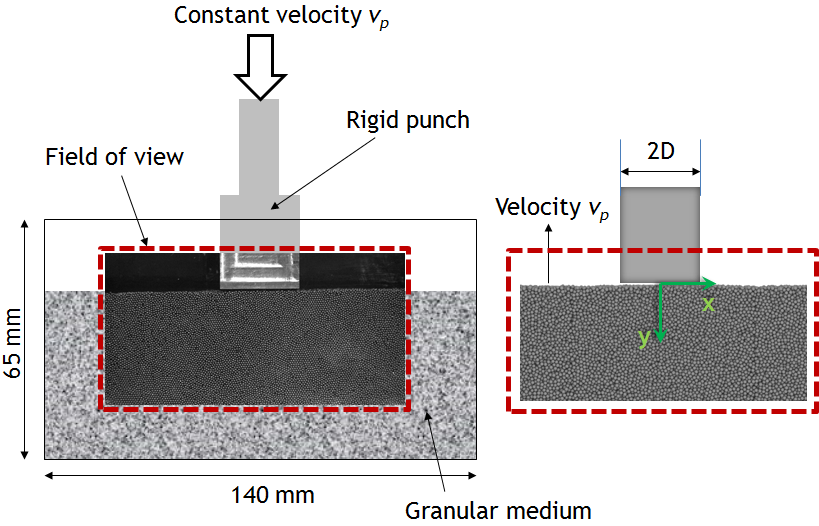}}
  }
  \\
  \mbox
  {
	\subfigure[\label{fig:3Dschematic}]{\includegraphics[scale=0.35]{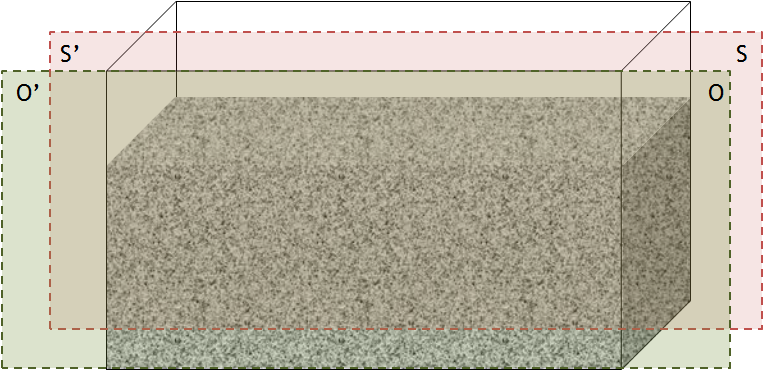}}
  }
  \caption{Problem setup. (a) (Left) Schematic of the experimental setup with sample image frame superimposed over the field of view (highlighted in red). (Right) Reference frame with punch stationary, showing the coordinate system used. The punch width $2D$, used for normalizing lengths, is also depicted. \notethis{(b) Perspective view of the container with the granular material. $O O^\prime$ is the plane of observation. Numerical simulations are performed for the entire 3D domain, including the front and back walls, corresponding to physical dimensions. Simulation results are visualized only for the center plane $SS^\prime$.}}
  \label{fig:expt_schematic}
\end{figure}

Figure~\ref{fig:expt_schematic} shows a schematic of the plane--strain indentation, wherein a flat--ended rectangular punch, made of stainless steel with dimensions 35 mm $\times$ 53 mm $\times$ 25 mm ($2D \times L_p \times T_p$), slowly penetrates the granular medium. The granular material consists of monodisperse spherical steel balls with diameter $d = 1$ mm. These were chosen for ease of tracking and to avoid the influence of size variation and particle deformation on the observed flow pattern. The balls were placed in a rectangular Al alloy box of dimensions 140 mm $\times$ 65 mm $\times$ 25 mm, as shown in the figure. The front face was covered with a glass plate to enable direct observation and imaging. 

The thickness $T_p$ of the punch (in the $z$-direction) matched that of the material, confining the punch motion to the $xy$ plane. The tolerance $\Delta T_p$ in the thickness dimension was maintained at $\pm 0.1$ mm ($\Delta T_p \ll d$), thereby ensuring that particles did not get into the gap between the punch wall and the front face of the box. The origin of coordinates was fixed at the midpoint of the bottom (flat) face of the punch, with the $x$-axis coinciding with the bottom face, i.e. along the width $2D$ of the punch --- see Fig.~\ref{fig:expt_schematic} (right).

The deforming granular medium  was imaged using a high--speed CMOS camera (PCO dimax), with a spatial resolution of $60$ $\mu$m per pixel. The field of view covered an area of 120 mm $\times$ 50 mm, and was illuminated by a 120 W halogen lamp light source. The experiments were done on a Universal Testing Machine (UTM) under velocity control. Three indentation velocities $v_p$ were used --- 0.02 mm/s, 0.2 mm/s and 2.0 mm/s --- covering 3 different orders of magnitude. Quantitative information about the flow field was obtained from the recorded images, using image analysis techniques. This involved first identifying the particles composing the medium in each frame, and then tracking them using Particle Tracking Velocimetry (PTV) methods. Post processing of experimental data is described in Sec.~\ref{subsec:postProcessing}.

\subsection{Numerical simulation}

Discrete Element Method (DEM) simulation of the punch indentation was performed to calculate the 3D kinematic properties of the medium. Several variants of DEM exist, each trading conceptual simplicity for stability and timestep length \cite{PoschelSchwager_CompGranDynamics,CundallStrack_Geotechnique_1979,CambouETAL_MicromechanicsGranular}. The dense granular material used in the experiments consisted of hundreds of thousands of persistent contacts. For this reason, we used the Non--Smooth Contact Dynamics (NSCD) method \cite{Jean_CompMethApplMechEngg_1999,Radjai_MechMat_2009} with the open source solfec code \cite{KoziaraBicanic_IntJNumMethEngg_2011} to carry out the simulations.

The generalized coordinates (positions and orientations) and velocities (linear and angular) of all particles are assembled in the vectors $\mathbf{q}, \mathbf{u}$. The complete equations for time--iteration can be represented as follows
\begin{align}
    \label{eqn:timestep1}
    \mathbf{q}^{t + \frac{h}{2}} &= \mathbf{q}^t + \frac{h}{2} \mathbf{u}^t\\
    \label{eqn:timestep2}
    \mathbf{u}^{t+h} &= \mathbf{u}^t + \mathbf{M}^{-1}\left( h \mathbf{f}^{t+\frac{h}{2}} + \mathbf{H}^T h \Lambda^{t+ \frac{h}{2}}\right)\\
    \label{eqn:timestep3}
    \mathbf{q}^{t+h} &= \mathbf{q}^{t+\frac{h}{2}} + \frac{h}{2} \mathbf{u}^{t+h}
\end{align}
with the superscripts denoting the values of the variables at timestep $t$, $t+\frac{h}{2}$ and $t+h$, with step size $h$. Here $\mathbf{M}$ is an inertia operator, $\mathbf{f}$ is a vector containing all external forces (such as gravity) and $\Lambda$ is the vector of reaction forces due to contact between particles. $\mathbf{H}$ is an operator that translates between the generalized coordinates and local contact coordinates \cite{Koziara_PhD_2008,KoziaraBicanic_IntJNumMethEngg_2011}. 

The basic idea behind the NSCD scheme is two-fold. Firstly, timestepping is done from $t$ to $t+h/2$, as in Eq.~\ref{eqn:timestep1}. Secondly, the coupled rigid-body constraint equations of the form
\begin{equation}
    \label{eqn:constraints}
    \mathbf{C}(\mathbf{H}\, \mathbf{u}^{t+h}, \Lambda) = 0
\end{equation}
are solved for $\Lambda$. The constraint equation arises from both impenetrability of particles and static/ dynamic friction (Signorini and Coulomb constraints respectively). Eq.~\ref{eqn:constraints} is solved after constraints are updated and new contact points detected, using the new positions $\mathbf{q}^{t + h/2}$, following Eq.~\ref{eqn:timestep1}. Timestepping is then continued, by updating the velocities in Eq.~\ref{eqn:timestep2} and positions from $t+h/2$ to $t+h$ in Eq.~\ref{eqn:timestep3}. This scheme enables the resolution of a large rigid--body system via a stable and efficient implicit solution, thus allowing large timesteps $h$. This approach is what differentiates the NSCD method from other DEM variants : it does not assume a Hertzian or spring contact model \cite{PoschelSchwager_CompGranDynamics} to compute reaction forces. For more details on the implementation, see Ref.~\cite{KoziaraBicanic_IntJNumMethEngg_2011}.

Dimensions in the simulations were chosen to correspond $1:1$ with the experimental setup, \notethis{shown in Fig.~\ref{fig:3Dschematic}}. The simulated granular medium was composed of 233,069 rigid spheres. The initial packing was formed using a collision driven packing algorithm \cite{SkogeETAL_PhysRevE_2006}, which generated a disordered packing of monodisperse spheres inside a cube of unit dimensions. Three such packings were assembled into a 3D box, with dimensions as in the experiments. The faces of the box were composed of rough rigid planes. This final ensemble was then allowed to relax under gravity using the NSCD method. A punch, also composed of rough rigid planar faces was then placed at the center of the relaxed sample and lowered at a constant velocity, maintained after every time step. This was done to mimic the velocity control used in the experiments. 

The $1:1$ simulation--to--experiment correspondence prevented a na\"{\i}ve translation of physical problem dimensions, which required very small timesteps, thereby increasing computation time. Hence, length and time were rescaled in the simulations, to ensure numerical stability during timestepping. Physical length $l_0$ and time $t_0$ were rescaled as $\hat{l} = \xi l_0$ and $\hat{t} = \chi t_0$ respectively with $\xi = 10^{3}, \chi = 31.645$. This ensured that the acceleration due to gravity was $\hat{g} = 10 \hat{l}/\hat{t}^2 = 10 l_0/t_0^2$. The simulation punch velocity $\hat{v} = \zeta v_0$ with $\zeta = 31.62$. The coefficient of friction $\mu$ values for all contacting surfaces --- ball--ball, ball--punch and ball--wall --- were set at $\mu = 0.4$, typical for steel contacts. \notethis{The visualized results are presented for the midplane of the simulation domain, see Fig.~\ref{fig:3Dschematic}}. 

\subsection{Post--processing}
\label{subsec:postProcessing}

The experimental data consisted of a sequence of high--resolution images of the medium flowing past the punch. The punch velocity $v_p$ was kept constant for each experiment, and the images were corrected to keep the punch stationary --- see Fig.~\ref{fig:schematic} (right). The grayscale images from the camera were represented by 2D floating point arrays, containing intensity data in the range $[0,1]$. Each frame was then processed to obtain the positions of the sphere centers and their displacements between successive frames, as described in Ref.~\cite{CrockerGrier_JCollIntSci_1996}. Since the frame rate was kept constant, velocities were obtained from inter--frame particle displacements over 15 frames. For particle $i$, the $x$ and $y$ components of the velocity were $u_i$ and $v_i$, $i = 1, 2, \cdots, N$ for $N$ tracked particles. The positions of the particles themselves were $(x_i, y_i)$. The scalars $u_i$ and $v_i$ represented two fields on a 2D plane, defined on the set of points with coordinates $(x_i, y_i)$. 

In the simulations, the positions $(x_i, y_i, z_i)$ and velocities $(u_i,v_i, w_i)$ of all the particles were recorded after every timestep. Deformation measures were then computed as necessary. Renders of particle positions were done using the Open Visualization Tool \cite{Stukowski_ModellSimMatSciEngg_2010}. 

\subsubsection{Deformation rate measures}

\begin{figure}
    \centering
    \includegraphics[scale=0.4]{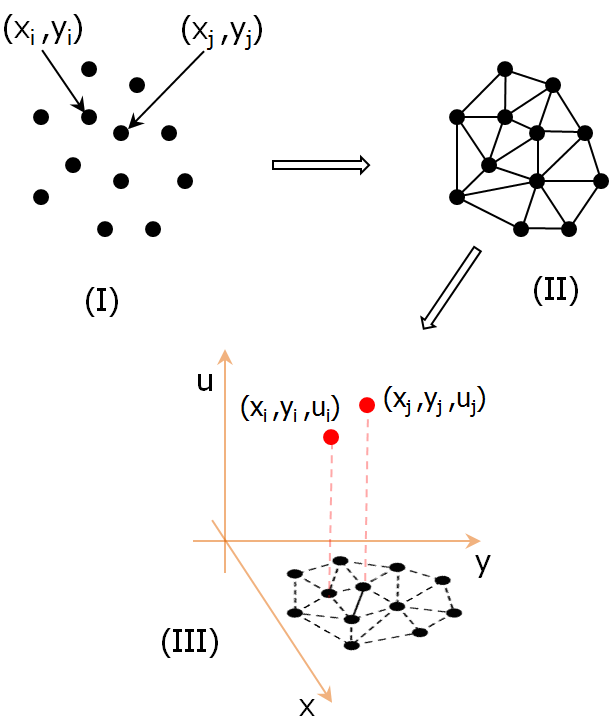}
    \caption{\label{fig:strainCalc} \notethis{Calculation of incremental strain measures from position and displacement data. The particle center positions and inter--frame displacements are stored (I). The position information $(x_i,y_i)$ is used to create a Delaunay triangulation of the entire ensemble (II). The 2D triangulation is converted to a 3D mesh by concatenating the particle's displacement (shown here for $u$, a similar procedure is used for $v$) to its coordinates (III). The resulting 3D mesh is used to approximate the $x$ and $y$ gradients of $u$ at each particle location.}}
\end{figure}

In order to determine various kinematic flow properties, an evaluation of the 2D velocity gradients $\nabla u \equiv (\partial u/\partial x, \partial u/ \partial y)$ and $\nabla v \equiv (\partial v/\partial x, \partial v/\partial y)$ was necessary. For a true continuum, these are usually approximated using finite differences on a uniform grid. However, given the discrete particulate nature of the medium, a true gradient (in the continuum sense) is not well-defined for a disordered grain network. Instead, velocity gradients were approximated as follows (\notethis{See Fig.~\ref{fig:strainCalc}}, the discussion is for $\nabla u$; the same applies to $\nabla v$). 
\begin{enumerate}
    \item A 2D Delaunay traingulation was performed with the obtained particle coordinates $(x_i, y_i)$ (\notethis{Fig.~\ref{fig:strainCalc}, parts (I) and (II)})
    \item The triangulation, with the scalar field $u_i$ determined a 3D heightfield mesh with vertices at $(x_i, y_i, u_i)$, corresponding to each tracked particle $i$. Implicitly, in 3D, this heightfield surface was given by $F(x,y) \equiv z - u(x,y) = 0$ (\notethis{Fig.~\ref{fig:strainCalc} (III)}).
    \item The vertex normals were calculated for each triangle of the 3D heightfield mesh, using methods of standard computational geometry \cite{DeBergETAL_ComputationalGeometry}. They were normalized to have unit norm and were of the form $(n_i^x, n_i^y, n_i^z)$.
    \item The heightfield, being of the form $F(x,y) = z - u(x,y) = 0$, had normals $\nabla F (x,y,z)$, given by 
\begin{equation}
\nabla F(x,y,z) = (-\partial u/\partial x, -\partial u/ \partial y, 1)
\end{equation}
So for discrete points $i$, $(\partial u/ \partial x)_i = -n_i^x/n_i^z$ and $(\partial u/ \partial y)_i = -n_i^y/n_i^z$ (when $n_i^z \neq 0$) approximated $\nabla u$ at the tracked particle $i$.
\end{enumerate}
The velocity gradients $\nabla u$ and $\nabla v$ at each point $i$ were then used to approximate continuum rate measures for the medium. 

The vorticity $\omega_i$ at each particle location, was calculated as:
\begin{equation}
\label{eqn:vorticity}
  \omega_i = \left(\frac{\partial u}{\partial y}\right)_i - \left(\frac{\partial v}{\partial x}\right)_i
\end{equation}

The strain rate tensor components were approximated by
\begin{align}
  \dot{\epsilon}_{xx} = \left(\pd{u}{x}\right)_i\quad \dot{\epsilon}_{yy} = \left(\pd{v}{y}\right)_i \\
\notag \dot{\epsilon}_{xy} = \frac{1}{2}\left[\left(\pd{u}{y}\right)_i + \left(\pd{v}{x}\right)_i\right]
\end{align}

From these relations, the 2D strain rate invariant was calculated for each particle $i$ for a particular frame as
\begin{equation}
\label{eqn:strainRateInvariants}
  \dot{\epsilon}_i^{S} = \frac{2}{3}\sqrt{\frac{1}{2}\left[\left(\dot{\epsilon}_{xx} - \dot{\epsilon}_{yy}\right)^2 + \dot{\epsilon}_{xx}^2 + \dot{\epsilon}_{yy}^2\right] + 3\dot{\epsilon}_{xy}^2}
\end{equation}

This provides a measure of the incremental shear strain in the medium. 

\subsubsection{Accumulated strain}

The strain rate tensor and vorticity were calculated from gradients of the instantaneous velocities. However, finite particle displacements during deformation invalidated the small displacement gradient approximation for the strain tensor. Due to temporal averaging used to calculate the particle displacements, integration of the incremental strain rate blurred the details of the strain field. Instead, the accumulated strain in the material was computed as follows (for more details, see Refs.~\cite{FalkLanger_PhysRevE_1998,ShimizuETAL_MatTrans_2007}).

For each particle $i$, the distance $\mathbf{l}_{ij}^t$ from its nearest neighbors $j$ was calculated after every timestep $t$. The deformation gradient $\mathbf{F}_i$ for particle $i$ was then determined by minimizing $\sum_j |\mathbf{l}_{ij}^0 \mathbf{F}_i - \mathbf{l}_{ij}^t|^2$, $\mathbf{l}_{ij}^0$ being the separation in the initial configuration. The Lagrangian strain tensor was defined as $\mathbf{E}_i = (\mathbf{F}_i \mathbf{F}_i^T - \mathbf{I})/2$. While the particles themselves were rigid, $\mathbf{E}_i$ measured material deformation resulting from relative motion of neighboring particles. 

From $\mathbf{E}$, the second shear strain invariant $\eta_i$ (in 3D) was calculated as
\begin{align}
\label{eqn:shearStrainInvariant}
  \eta_i = \bigg[&E_{yz}^2 + E_{xz}^2 + E_{xy}^2 + \\
&\frac{(E_{yy} - E_{zz})^2 + (E_{xx} - E_{zz})^2 + (E_{yy} - E_{xx})^2}{6}\bigg]^{\frac{1}{2}}
\end{align}
and provided a scalar measure of the material shear strain, corresponding to the particle $i$. It has been shown that $\eta_i$ captures strain due to local structural rearrangements \cite{FalkLanger_PhysRevE_1998,ShimizuETAL_MatTrans_2007}. For the 2D case, $E_{xz} = E_{yz} = E_{zz} = 0$ in the relation for $\eta_i$ above.

The expressions for  $\epdt{S}$ and $\eta_i$ in terms of components of the strain rate ($\dot{\epsilon}_{ab}$)and strain tensors ($E_{ab}$) are identical but their methods of computation are different.


\section{Results}
\label{sec:results}
The individual particle trajectories $(x_i(t), y_i(t))$, velocities $(u_i, v_i)$ and gradients $(\nabla u)_i, (\nabla v)_i$ (in the plane $xy$ of observation) were extracted from the high--resolution image sequence (\emph{cf.} Sec.~\ref{subsec:postProcessing}). Simultaneously, 3D particle positions and velocities were obtained from $1:1$ numerical NSCD simulations. Movie M1 (see supplemental material \cite{SuppMat}) shows a typical sequence of images from an experiment and the corresponding simulation.

\subsection{Particle tracking in plane--strain}

\begin{figure}
\clearpage
  \centering
  \mbox
  {
    \subfigure[\label{fig:planeStrain}]{\includegraphics[scale=0.45]{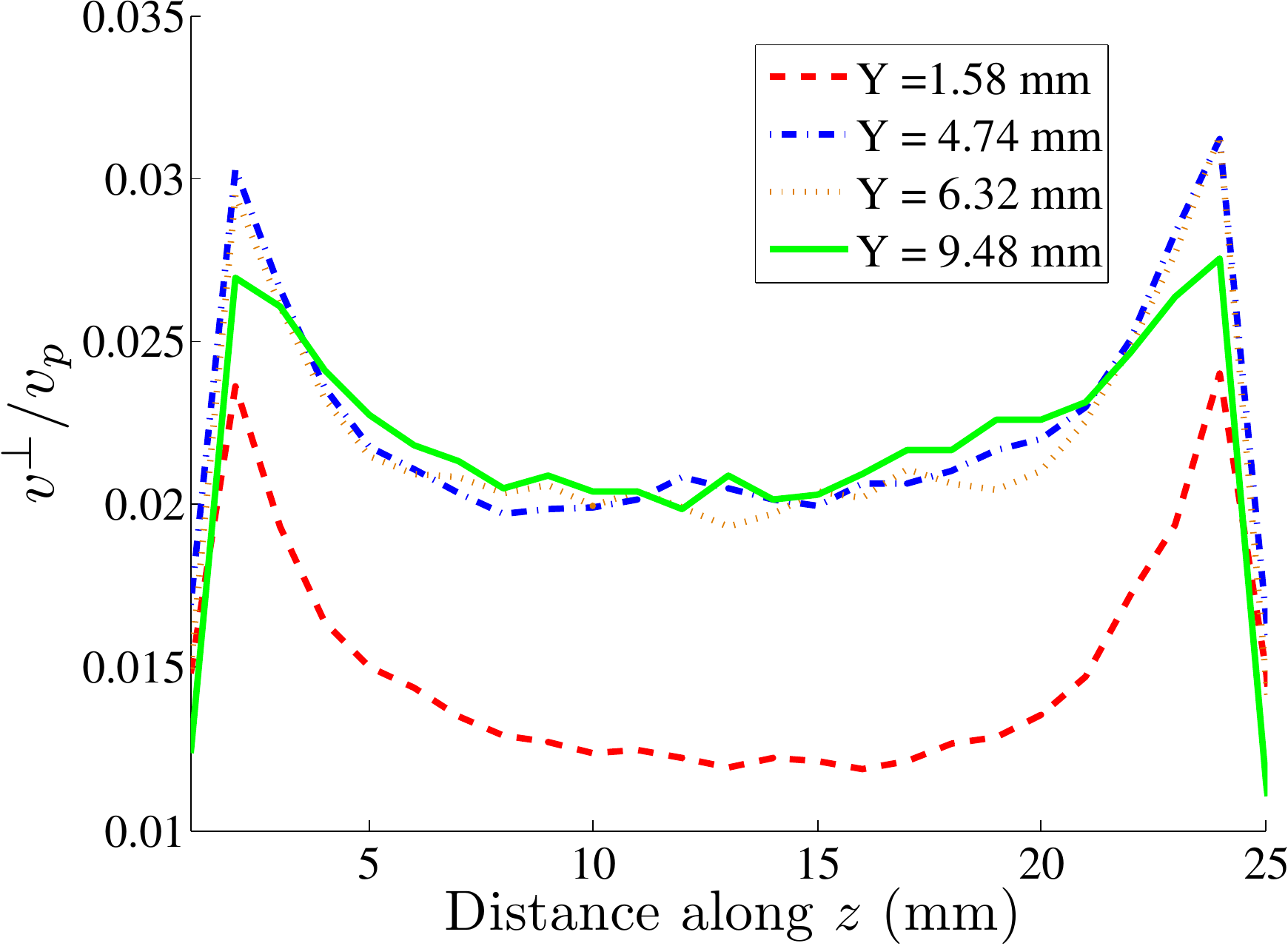}}
  }
  \\
  \mbox
  {
    \subfigure[\label{fig:tracks}]{\includegraphics[scale=0.45]{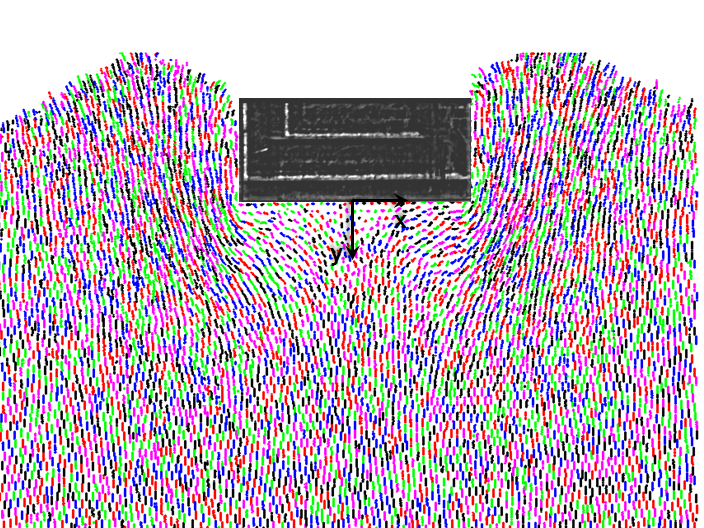}}
  }
  \caption{Validity of plane--strain assumption at the mesoscopic scale. (a) Plot of particle velocity $v^\perp$ normal to observation plane in a simulation run. This velocity decreases quite rapidly with distance from the front and back faces of the container box. (b) Colored particle paths from successive experimental images. Punch velocity $v_p = 0.2$ mm/s.}
  \label{fig:planeStrainValidity}
\end{figure}

Macroscopically, the punch is constrained to move in the $xy$-plane. However, microscopically, the particles composing the granular material are free to move in all three directions $x, y$ and $z$. For media such as sand, where the particle size is much smaller than a typical macroscopic length scale, the plane strain assumption suffices to describe the flow field accurately \cite{Nedderman_StaticsKinematicsGranular}. An analysis of the particle velocities obtained from the simulations was used to verify the applicability of the plane flow approximation in the current experiment.

For this purpose, particle velocities $v^\perp (\equiv |v_z|$, see coordinate system in Fig.~\ref{fig:expt_schematic}), perpendicular to the viewing ($xy$) plane, are extracted from the NSCD simulations, for parallel planes with normal along the $z$-axis. Figure \ref{fig:planeStrain} shows the variation of the average $v^\perp/v_p$ in each such plane, for increasing punch penetration depth $Y$. The graph is quantitatively similar for the different $v_p$ considered. The average value of $v^{\perp}/ v_p$ is very small at the two peaks ($\sim 3.5 \%$), which occur just inside the front and back faces of the sample. Moreover, after a short initial rise (between $Y = 1.58$ mm and $Y = 3.16$ mm),  $v^\perp$ remains unchanged. The simulations thus indicate that the net contribution to flow in the $z$-direction, perpendicular to the viewing plane, is very small. This justifies the use of particle tracking methods restricted to the $xy$ plane.

A typical set of tracks, obtained from an analysis of the experimental data for $v_p = 0.2$ mm/s, is displayed in Fig.~\ref{fig:tracks}. Each color line represents the trajectory of the center of an individual particle in the medium, over 30 successive frames. Particle tracking provided $(x_i, y_i)$ and $(u_i, v_i)$, corresponding to particle center coordinates and inter--frame displacements (i.e. instantaneous velocities) respectively. The velocity gradients were approximated from this data (Sec.~\ref{subsec:postProcessing}). 

\subsection{Stagnation zone and its evolution}
\label{subsec:stagnationZone}
\begin{figure}
\clearpage
  \centering
  \mbox
  {
    \subfigure[\label{fig:flowFeatures}]{\includegraphics[scale=0.4]{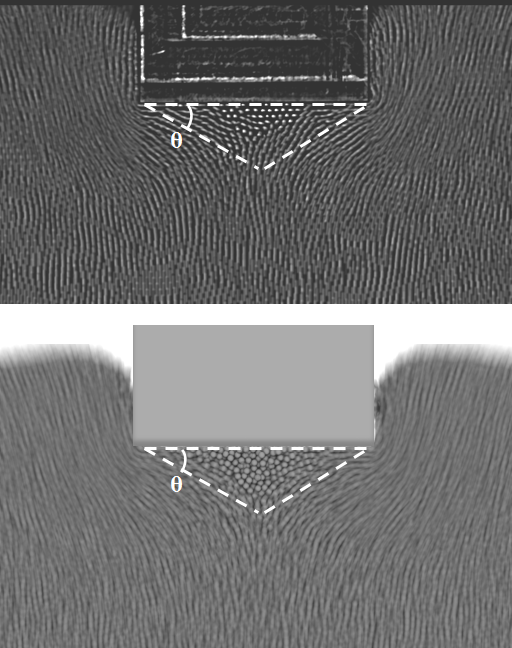}}
  }
  \\
  \mbox
  {
    \subfigure[\label{fig:angleTimePlot}]{\includegraphics[scale=0.4]{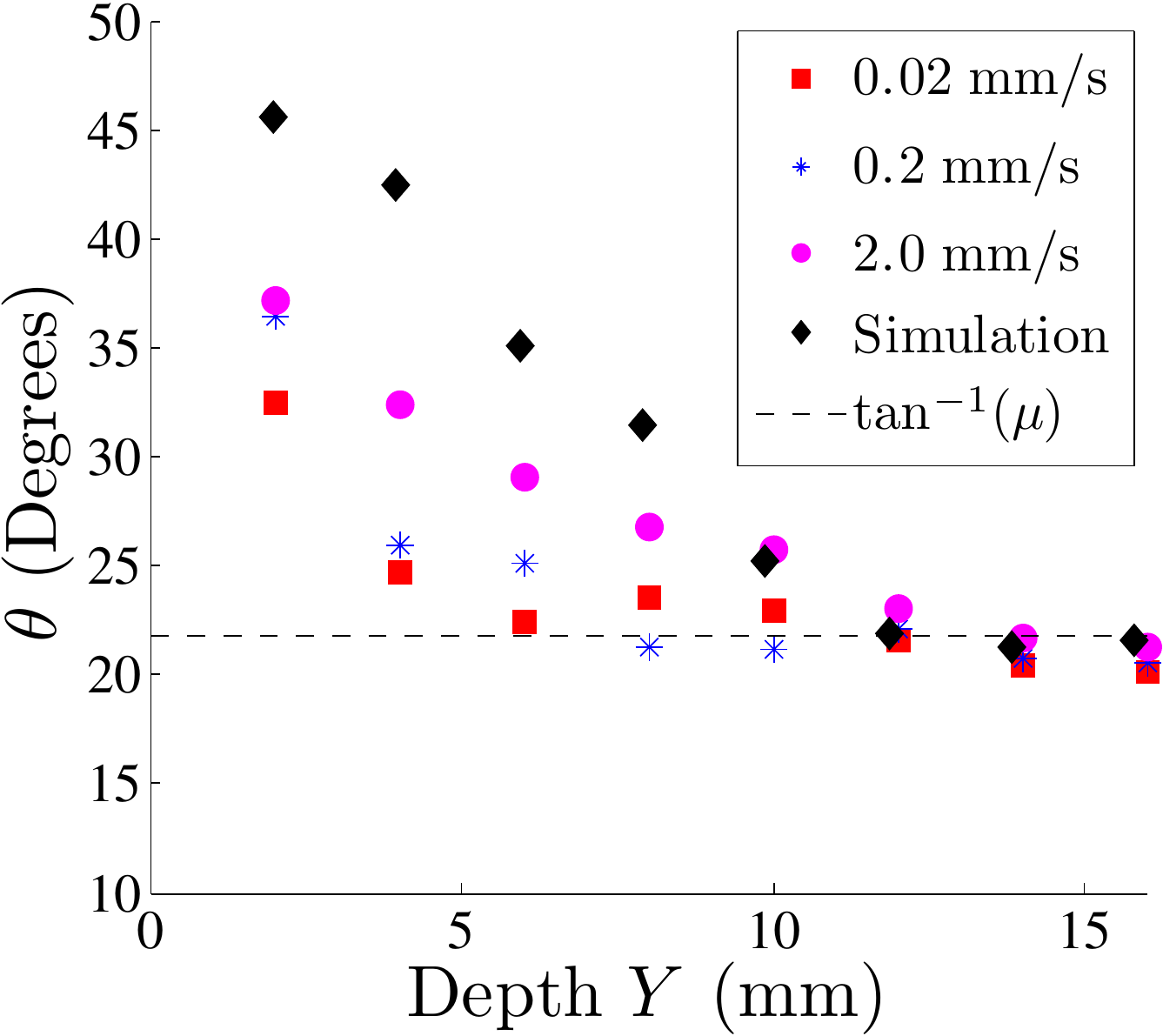}}
  }
  \caption{Stagnation (or dead material) zone and its evolution. (a) Sequence of 100 experimental (top) and simulation (bottom) frames superimposed, showing the flow pattern. The streaks correspond to particle pathlines. The stagnation zone (white dashed line) appears motionless, as revealed by lack of streaklines. (b) Variation of base angle ($\theta$) of the stagnation zone with punch penetration depth $Y$. The dashed line represents $\tan^{-1}(\mu)$ used in the simulations.}
  \label{fig:deadMaterialZoneEvolution}
\end{figure}

Keeping the punch stationary, 100 experimental images are stacked on top of each other to reveal the flow of individual grains, as in Fig.~\ref{fig:flowFeatures} (top). Likewise, equivalent NSCD simulation data is rendered into images corresponding to successive timesteps, and the images then stacked to obtain the bottom image shown in Fig.~\ref{fig:flowFeatures}. The two images are very similar, consistent with the physical conditions being the same. This can also be seen from Movie M1 \cite{SuppMat}, which provides a visual comparison between experiment and simulation. The grains flow around the punch, creating tracks --- particle pathlines --- clearly seen in both composite images.

A characteristic feature of the flow is the occurrence of a stagnation zone, or region of \lq dead material\rq , underneath the punch. This region consists of particles that are stationary relative to the punch. The zone is roughly triangular in shape, as seen in Fig.~\ref{fig:deadMaterialZoneEvolution}, and symmetrical with respect to the $y$-axis, consistent with the loading condition. The angle $\theta$ of the base of the triangle provides a measure of the shape of the zone, see Fig.~\ref{fig:flowFeatures}. \notethis{In order to measure the evolution of $\theta$, different composite images (indexed by $k$) are generated by combining 100 images around a particular value of $Y$. The top left and top right corners as well as the lowest point of the stagnation zone are determined for each composite image. These correspond to the centers of the leftmost, rightmost and bottom-most stationary particles in the composite image respectively. Two lines are then drawn joining the lowest point with the top corners; the angle subtended is $\theta_k$. $\theta_k$ values are averaged ($k=10$) to give $\theta$ for the particular $Y$ value in Fig.~\ref{fig:deadMaterialZoneEvolution}.}

$\theta$ changes with the punch penetration depth $Y$, but the stagnation zone remains nearly triangular and symmetric. A plot of $\theta$ vs. $Y$, as measured from experimental data, is shown in Fig.~\ref{fig:angleTimePlot}. Also shown is the corresponding $\theta$ value obtained from numerical simulations, for $v_p = 0.5$ mm/s. The simulations accurately reproduce the experimental stagnation zone geometry. The initial value $\theta \sim 50^\circ$  was found to be sensitive to the packing details, as confirmed by observing data from repeated simulation runs. \notethis{After indentation to a sufficient depth, $\theta$ tends to a limiting value $\simeq \tan^{-1} \mu$, independent of the sliding velocity. This behavior is also reproduced by the NSCD simulations. Here $\mu = 0.4$ is the friction coefficient between all the surfaces in contact. We note that there was no observable flow along the bottom wall of the container for the $Y$ values reported in Fig.~\ref{fig:angleTimePlot}.}

\notethis{On a longer timescale, the size of the stagnation zone decreases slowly with $Y$, even though $\theta$ reaches a steady value. This is because friction causes a layer of particles immediately adjoining the face of the punch to move sideways. However, on the scale of a few ($\sim 5$) mm, the change in the stagnation zone width is negligible compared to $Y$, hence the particles do not form distinct pathlines in the composite image.}


\subsection{Rotation field and vorticity}
\label{subsec:vorticity}

\begin{figure}
\clearpage
  \centering
  \mbox
  {
    \subfigure[\label{fig:velFieldExp}]{\includegraphics[scale=0.38]{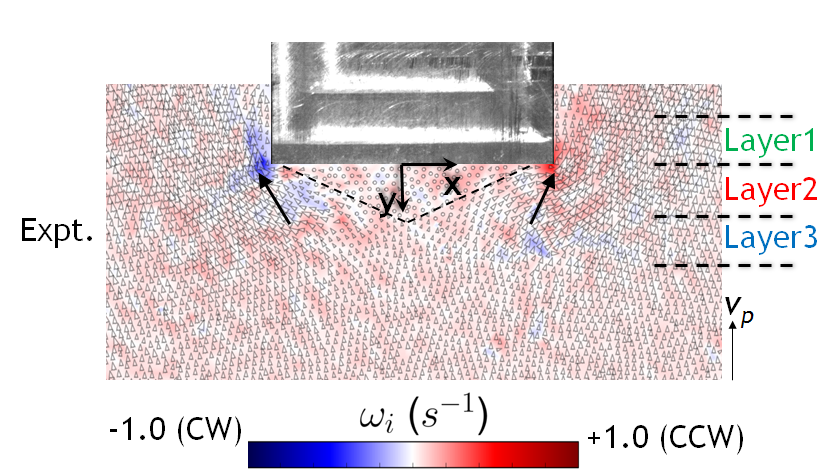}}
  }
  \\
  \mbox
  {
    \subfigure[\label{fig:velFieldSim}]{\includegraphics[scale=0.38]{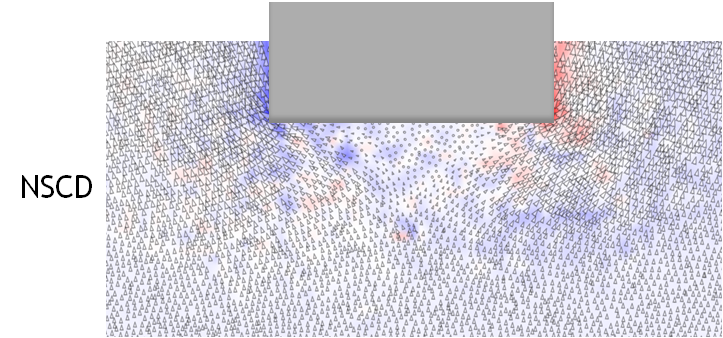}}
  }
  \caption{Particle velocities and vorticity field near the punch. Arrows represent velocity of individual particles, estimated from successive images; arrow size is proportional to velocity magnitude. Background is colored based on vorticity $\omega_i$ (Eq.~\ref{eqn:vorticity}) using linear interpolation from values at particle locations. $\omega_i$ is scaled between -1 (blue, clockwise) and 1 (red, counterclockwise). (a) Experimental image for $v_p = 0.2$ mm/s. Arrows point to the vortex centers, while stagnation zone is demarcated by black dashed lines. The layers used for vorticity plot are also indicated. (b) Corresponding image from NSCD simulation --- both the stagnation zone geometry and vortex centers are quantitatively reproduced in the simulations.}
  \label{fig:particleVelocityVorticity}
\end{figure}

\begin{figure}
\clearpage
    \centering
    \includegraphics[scale=0.25]{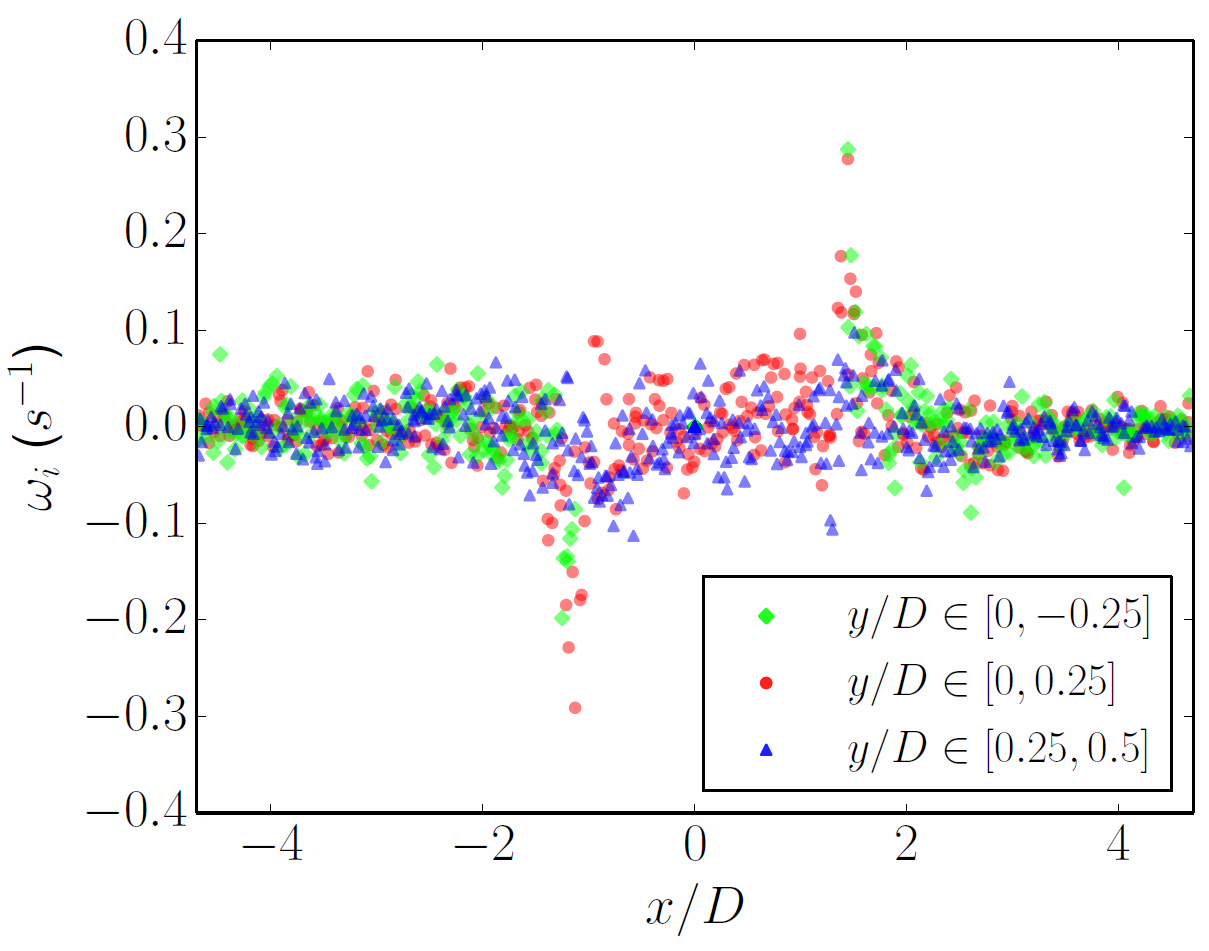}
    \caption{\label{fig:rotScatter} Scatter plot of rotation field approximated from experiments, in three horizontal layers adjoining the bottom face of the punch. The two peaks in the layer just below the punch show the occurence of vortices at $x/D \sim \pm 1.25$.}
\end{figure}


In the coordinate system fixed to the punch, flow separation occurs at the bottom face of the punch. The vorticity $\omega_i$ (Eq.~\ref{eqn:vorticity}) was used to quantify the net rotation of the medium, induced by differential displacements of adjacent particles. 

Figure~\ref{fig:particleVelocityVorticity} shows a plot of the vorticity, linearly interpolated from $\omega_i$ values at each particle location $i$. $\omega_i$ is scaled to lie between -1 (blue, clockwise) and +1 (red, counterclockwise). Individual particle velocities are superimposed on top using arrows. The arrow size reflects the velocity magnitude. Figure.~\ref{fig:velFieldExp} shows the result for an experiment with $v_p = 0.2$ mm/s. Qualitatively similar plots were obtained for the other $v_p$. For a comparison, an equivalent plot using simulation data is shown in Fig.~\ref{fig:velFieldSim}. In this latter case, particles in a narrow region (of width 1.5 particle diameters) were isolated at the center along the $z$-direction (\emph{cf}. Figs.~\ref{fig:3Dschematic} and ~\ref{fig:planeStrain}). 

The stagnation zone is marked by dashed lines in Fig.~\ref{fig:velFieldExp}. Far away from the punch, the material flows with velocity $v_p$, as indicated in the figure. Two centers of vorticity are observed, indicated by the two arrows, corresponding to regions with high $\omega_i$. These two vortices are of opposite orientation, due to flow separation at the punch bottom face. As the material approaches the bottom face, individual particles are forced to move away from the centerline $(x=0)$, thereby flowing around the side faces of the punch.

To better understand the variation in material rotation along the $y$ direction, $\omega_i$ from the experimental data in three neighboring spatially fixed layers (see Fig.~\ref{fig:velFieldExp}) is plotted in Fig.~\ref{fig:rotScatter}. These layers are of equal width and adjacent to the bottom face of the punch. The scatter values of $\omega_i$ in a layer immediately above the bottom face (Layer 1) are plotted in green. The values in a layer right below the punch (Layer 2) are shown in red. $\omega_i$ in the adjacent layer (Layer 3) are plotted in blue. In Layer 1, a sharp increase is seen in $\omega_i$, just around the lateral edges of the punch. This occurs at $x/D \simeq \pm 1.2$. In Layer 2, two peaks, corresponding to the vortices in Fig.~\ref{fig:velFieldExp}, are observed on either side of the punch. These peaks occur at $x/D \simeq \pm 1.25$, immediately below the punch bottom face. In Layer 3, the vorticity is significantly diminished. Hence, finite vorticity remains confined to a thin layer near the bottom face of the punch, decaying rapidly away from it.

The observed rotation field was also well reproduced in the NSCD simulations (Fig.~\ref{fig:velFieldSim}), including the two vortices on either side of the punch. As a result of interparticle friction, velocities differ substantially from $v_p$ only in a region near the punch. This is the cause for the rapid decrease in vorticity along the $y$-axis, in Fig.~\ref{fig:rotScatter}. Since the simulation data were extracted in the middle of the container, the vortex lines are parallel to the $z$-axis and extend throughout the thickness of the sample. Furthermore, it was observed that the vortices (in both experiments and simulations) were sustained even after the angle $\theta$ of the dead material zone reached its steady value (\emph{cf.} Fig.~\ref{fig:angleTimePlot}). The vortices hence exist during steady flow --- similar observations of vortices have been made in experiments with dry sand \cite{MurthyETAL_PhysRevE_2012}.

\subsection{Shear band formation}
\label{subsec:shearBandFormation}

\begin{figure}
\clearpage
  \centering
  \mbox
  {
    \subfigure[\label{fig:strainRate}]{\includegraphics[scale=0.43]{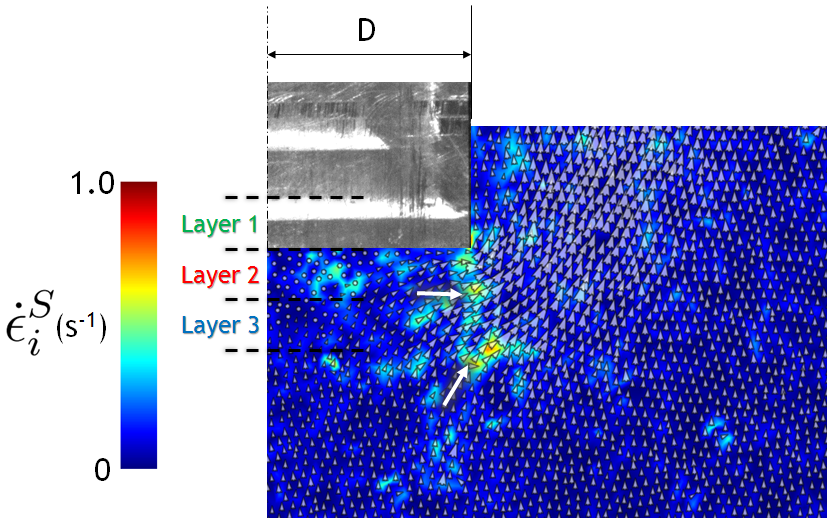}}
  }
  \\
  \mbox
  {
    \subfigure[\label{fig:epdtS}]{\includegraphics[scale=0.25]{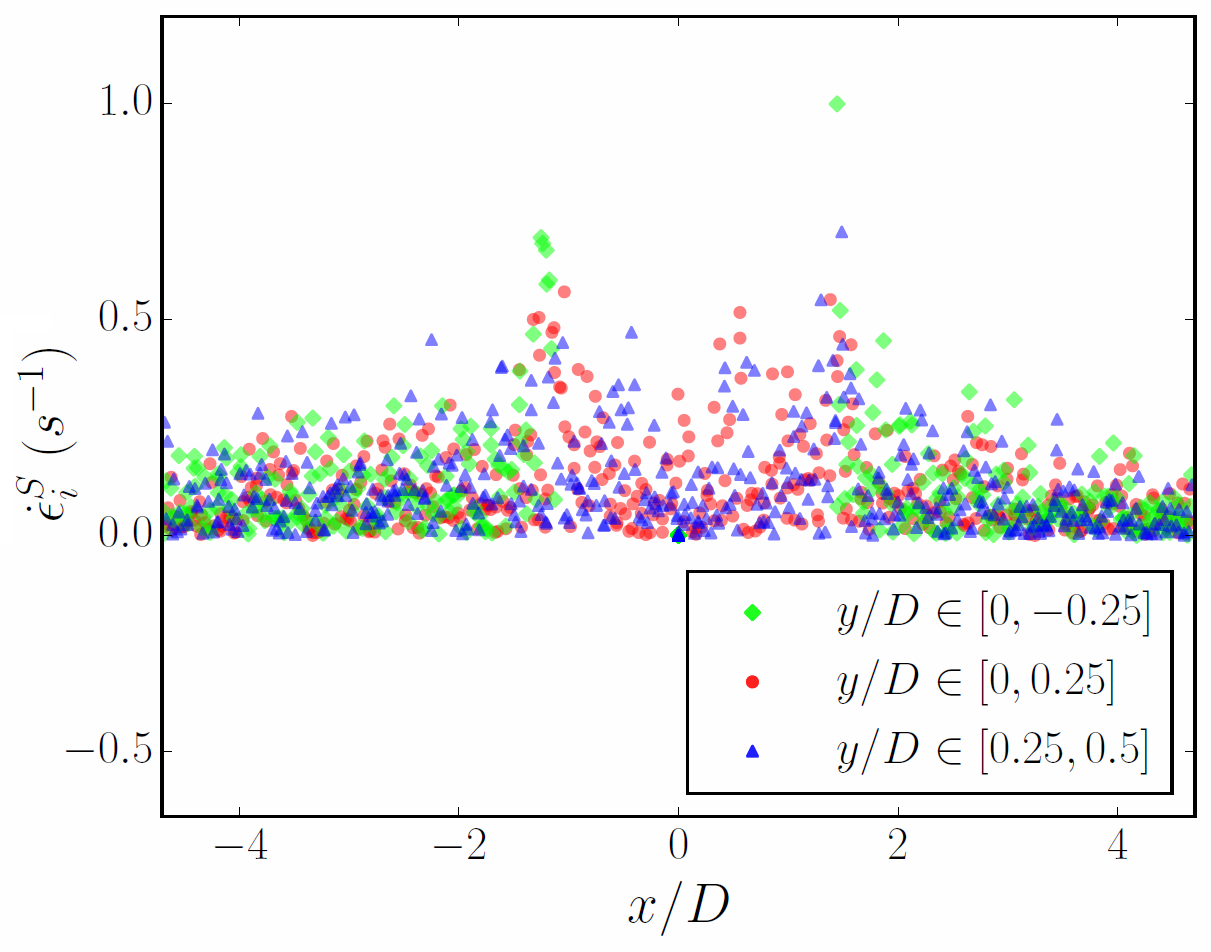}}
  }
  \caption{Material strain rate field from experimental data. (a) Half the (symmetric) flow field near the punch with particle velocities (arrows) superimposed on shear strain rate invariant $\dot{\epsilon}_i^S$ (Eq.~\ref{eqn:strainRateInvariants}). (b) Scatter plot of $\dot{\epsilon}_i^S$ in the three layers shown in (a). High strain rates accompany large changes in neighboring particle velocities. }
  \label{fig:strainRateVariation}
\end{figure}

It is well known that fine--grained granular media undergo shear as they flow around a punch \cite{Nedderman_StaticsKinematicsGranular,SchallvanHecke_AnnRevFluMech_2009}. Due to local rearrangements of particles, shear strain tends to get concentrated in the form of bands that demarcate regions with a sharp difference in the velocity of adjacent particles.

An estimate of velocity gradients resulting in shear was obtained from the invariant measure $\dot{\epsilon}_i^S$ (Eq.~\ref{eqn:strainRateInvariants}). The $\dot{\epsilon}_i^S$ field, linearly interpolated between particle locations, along with superimposed particle velocities, is shown in Fig.~\ref{fig:strainRate}. The two white arrows indicate regions of high shear rate, occuring right below the punch. These regions are seen to coincide with a sharp gradient in particle velocity magnitude. Moreover, the fluctuations in $\epdt{S}$, evident in Fig.~\ref{fig:strainRate}, reflect local particle rearrangements. 

A scatter plot of $\epdt{S}$, corresponding to the three layers depicted in Fig.~\ref{fig:strainRate}, is shown in Fig.~\ref{fig:epdtS}. In Layer 1, just above the bottom face of the punch (green), large shear strains are observed very close to the sides of the punch. In Layer 2 (red), high values of $\epdt{S}$ occur below the punch, due to particle rearrangements. Finally, significant shear rates are also seen in Layer 3 (blue). This is in contrast to the vorticity, which decreases rapidly in this layer (\emph{cf.} Fig.~\ref{fig:rotScatter}). 

\begin{figure}
\clearpage
  \centering
  \mbox
  {
    \subfigure[\label{fig:VMplotA}]{\includegraphics[scale=0.27]{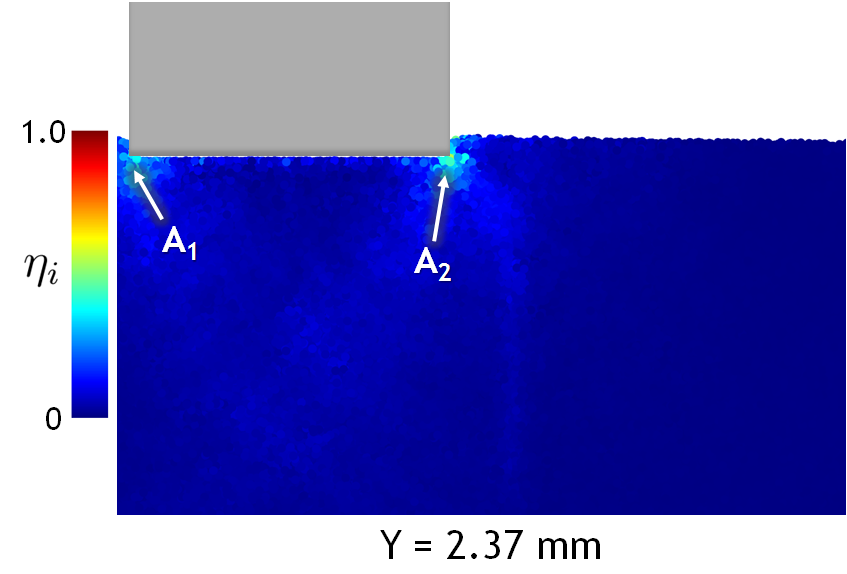}}
  }\\
  \mbox
  {
    \subfigure[\label{fig:VMplotB}]{\includegraphics[scale=0.27]{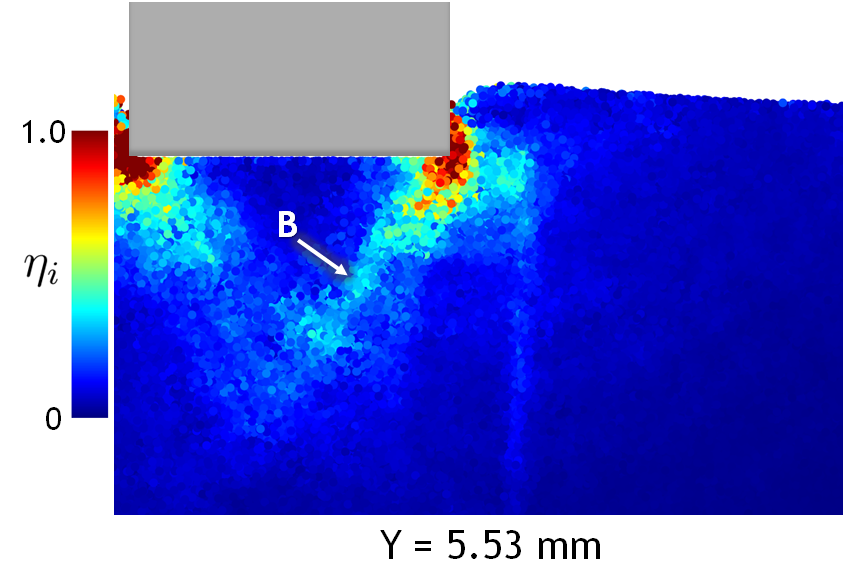}}
  }\\
  \mbox
  {
    \subfigure[\label{fig:VMplotC}]{\includegraphics[scale=0.27]{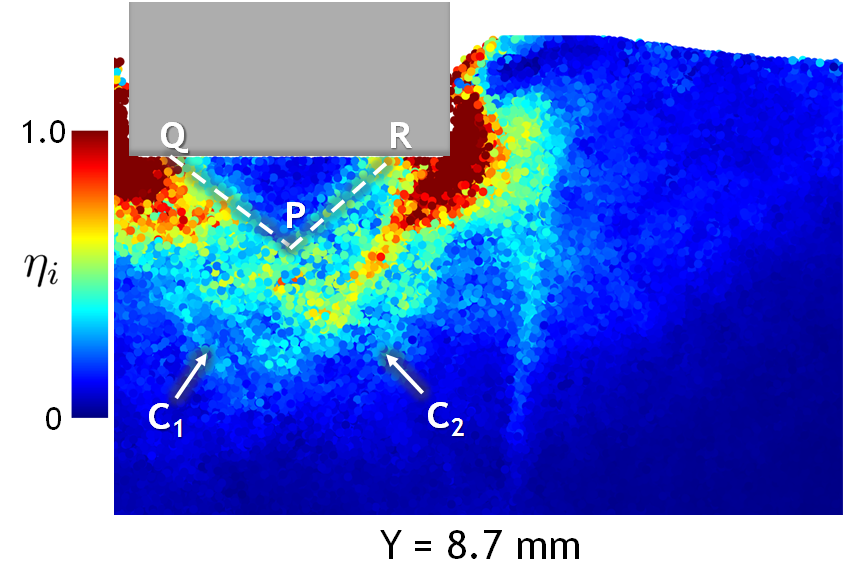}}
  }

  \caption{Calculated strain data from NSCD simulation for $v_p = 0.5$ mm/s; the particles are colored by the value of $\eta_i$. Only half of the flow field is shown. (a) $Y = 2.37$ mm. Strain accumulation begins at the two bottom edges $A_1$ and $A_2$ of the punch. (b) $Y = 5.53$ mm. A shear band $B$ is seen forming underneath the stagnation zone. (c) $Y = 8.7$ mm. The stagnation zone $PQR$ is clearly seen, surrounded by regions of shear in two bands. Additional shear bands $C_1$ and $C_2$ are beginning to develop below the stagnation zone.}
  \label{fig:strainAccumulation}
\end{figure}

Regions of high velocity gradient occur at nearly constant spatial locations, causing shear strain accumulation in the form of continuously evolving narrow bands. A quantitative measure of this strain is provided by the strain invariant $\eta_i$ (Eq.~\ref{eqn:shearStrainInvariant}). A rendered sequence from an NSCD simulation, with particles colored by their $\eta_i$ value, is shown in Fig.~\ref{fig:strainAccumulation}. Movie M2 \cite{SuppMat} shows a time sequence of the evolution of $\eta_i$ in the material. Each successive frame in Fig.~\ref{fig:strainAccumulation} corresponds to increasing punch penetration $Y$. Shear in the material first occurs at the two edges $A_1$ and $A_2$, near the bottom of the punch (Fig.~\ref{fig:VMplotA}). As the material continues to flow, the formation of two distinct shear bands ($B$) is seen in Fig.~\ref{fig:VMplotB}. The two bands are symmetric with respect to the vertical, and bound the incipient stagnation zone underneath the punch. The ends of these bands coincide with the edges $A_1$ and $A_2$, so that they run upto the bottom face of the punch. With increasing $Y$, shear strain in these bands further increases (Fig.~\ref{fig:VMplotC}). This causes additional shear bands to form underneath the stagnation zone, see points $C_1$ and $C_2$ in Fig.~\ref{fig:VMplotC}. At this stage, the stagnation zone geometry is near its steady state value (i.e. $\theta$ nearly constant). The total strain in this zone is much lesser than in the material immediately surrounding it, even though small local fluctuations in strain rate are observed --- as seen in Movie M2 and Fig.~\ref{fig:strainRateVariation}. 

\begin{figure}
\clearpage
  \centering
  \mbox
  {
    \subfigure[\label{fig:shearBandPos}]{\includegraphics[scale=0.42]{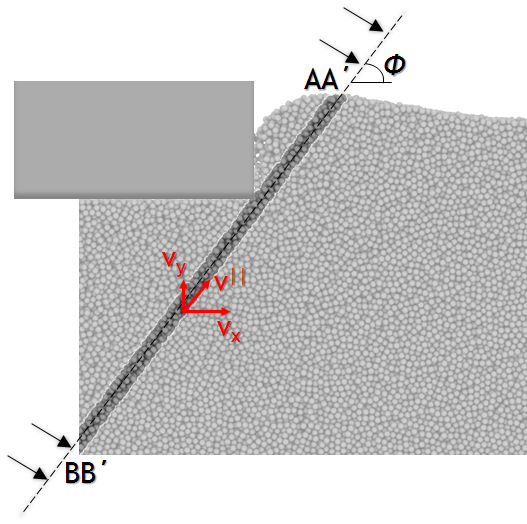}}
  }\\
  \mbox
  {
    \subfigure[\label{fig:shearBandVel}]{\includegraphics[scale=0.42]{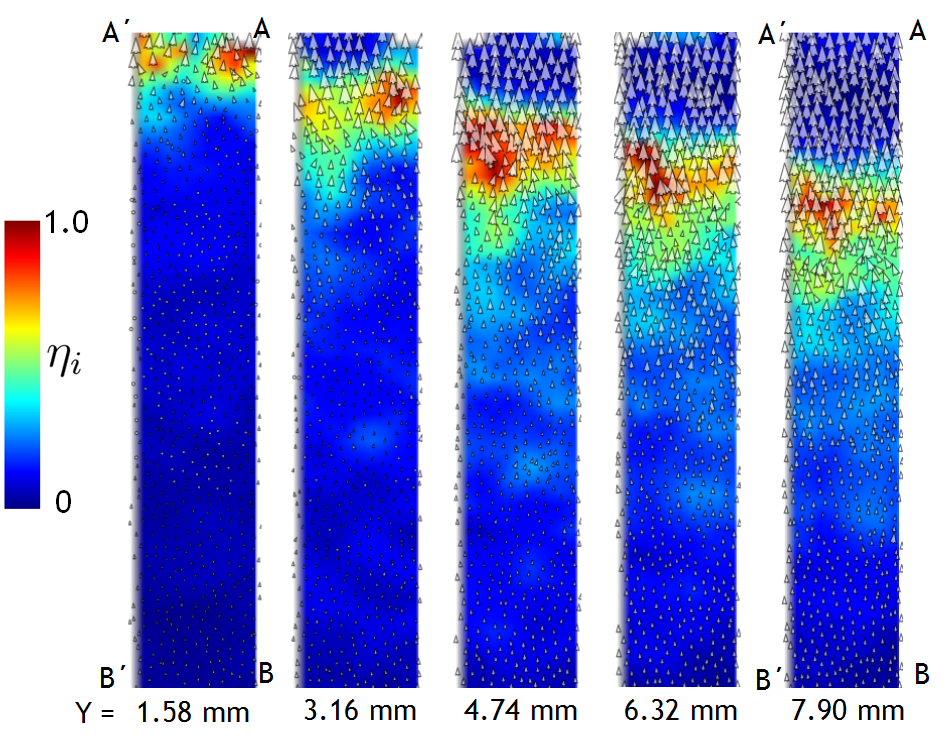}}
  }
  \caption{Build--up of shear in a narrow region. (a) Spatial location for tracking particle motion. The plane $AA^\prime BB^\prime$ passes through the thickness of the sample (in the $z$-direction). (b) Material in the plane $AA^\prime BB^\prime$, viewed along the plane normal. Arrows represent in--plane particle velocity $v^{||}$. The color depicts linearly interpolated strain invariant $\eta$ for the medium. Direction of particle motion indicates reduction in density along the band.}
  \label{fig:shearBandVelocities}
\end{figure}

These observations suggest that large shear strains are concomitant with steep velocity gradients in the flow. Further evidence is provided by observing particles in the plane of the shear band itself. To this end, particles at the spatial location of the band $B$ (Fig.~\ref{fig:VMplotB}), marked by the plane $AA^\prime BB^\prime$ in Fig.~\ref{fig:shearBandPos}, are isolated. This plane is chosen to subtend an angle $\phi = \arctan({4/3})$ with the $x$-axis. When viewed along the normal to this plane, the constituent particles form a thin rectangular layer, with width $AA^\prime$ equal to the thickness of the granular material. The in--plane velocity $v_i^{||}$  of each particle $i$ is calculated from the $x$ and $y$ velocities $v_x, v_y$ as
\begin{equation}
  v^{||} = v_x \cos\phi + v_y \sin\phi
\end{equation}
The velocities $v^{||}$ in the plane $AA^\prime BB^\prime$ are shown by arrows in Fig.~\ref{fig:shearBandVel}, with $Y$ increasing from left to right. The background color in each of these images is calculated from the strain invariant $\eta$ for the medium, obtained using linear interpolation from the particles' individual $\eta_i$ values. Initially, for small $Y$, a zone of shear develops near the bottom edges of the punch, close to the end $AA^\prime$. This corresponds to the material near $A_2$ in Fig.~\ref{fig:VMplotA}. As the plane $AA^\prime BB^\prime$ was chosen to be fixed in space, this zone of shear moves downward with increasing $Y$. The cause for this is the material pileup on either side of the punch (see sequence in Fig.~\ref{fig:strainAccumulation}). The zone of shear further increases in size with increasing $Y$, seen by its extent for $Y = 3.16$ mm and $Y = 7.90$ mm in Fig.~\ref{fig:shearBandVel}. 

Another interesting feature observed in Fig.~\ref{fig:shearBandVel} is the particle velocity distribution. While particles below the shear zone remain stationary, those above the zone continue moving upward; \emph{cf.} arrow sizes in Fig.~\ref{fig:shearBandVel} for $Y = 6.32$ mm and $Y = 7.90$ mm. The shear zone hence corresponds to high velocity gradients in the material, and low shear exists even when the particles in the material move with significant velocity. The motion of particles away from the end $BB^\prime$ of the plane $AA^\prime BB^\prime$ also results in significant local dilatancy in the material. Consequently, highly sheared regions result from irreversible motion of the consituent particles, in turn causing volume change in the material.

\subsection{Other structural changes}
In addition to the flow features discussed above, dense flow in the punch indentation configuration shows other interesting structural changes. Since these changes cannot be observed directly by 2D imaging, they are extracted from the 3D simulation data. Two such features are highlighted and discussed.

\subsubsection{Free surface evolution}
\label{subsubsec:freeSurface}
\begin{figure}
    \centering
    \includegraphics[scale=0.25]{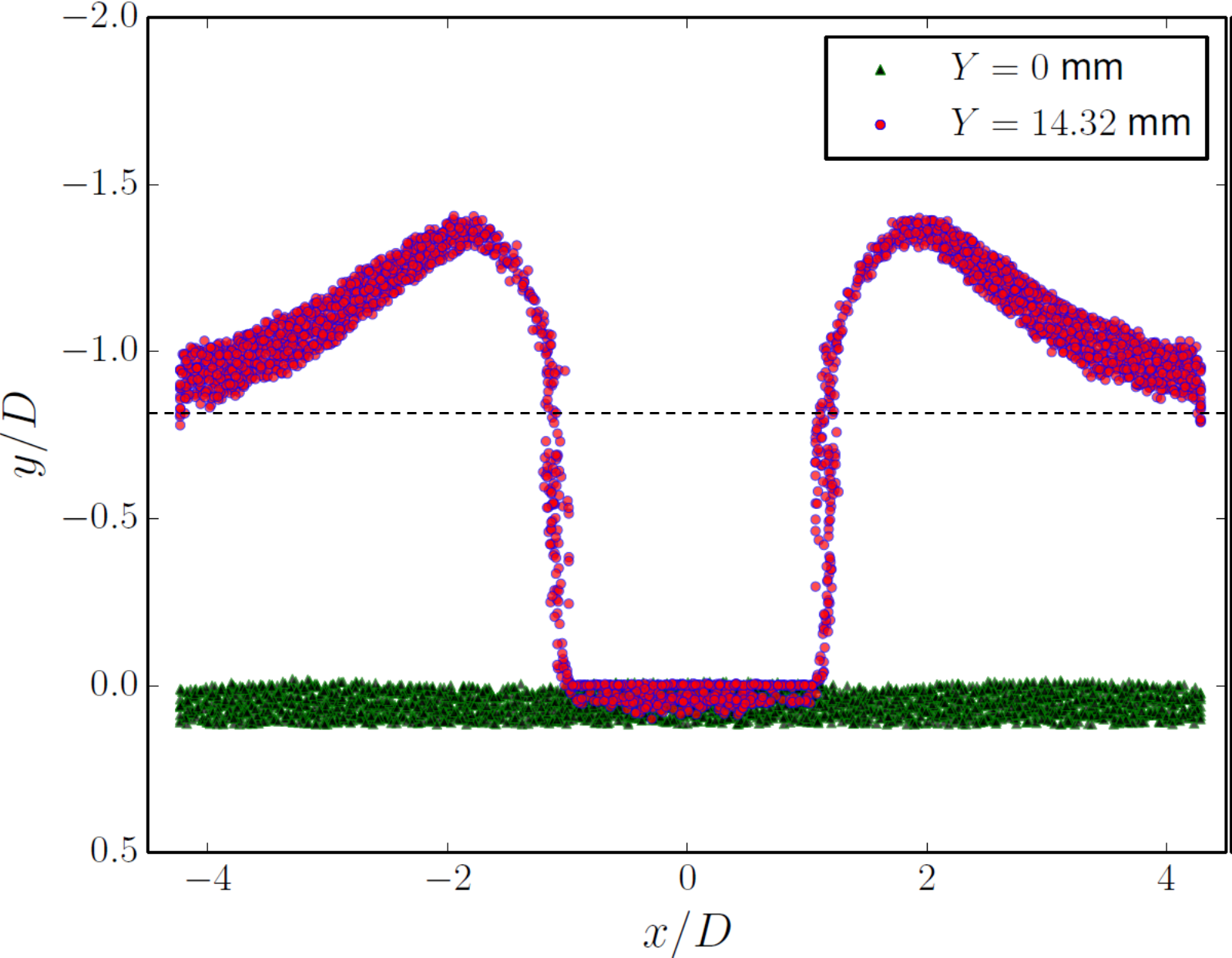}
    \caption{\label{fig:freeSurface} Initial ($Y=0$) and final ($Y=14.32$ mm) free surface profile of the medium. New surface formation along the sides of the punch occurs by movement of grains from the bulk. Dotted line denotes final free surface position in the absence of the punch.}
\end{figure}

In punch indentation of metals, it is known that new surfaces are created along the sides of the punch, with increasing penetration $Y$ \cite{SamuelsMulhearn_JMechPhysSolids_1957}. This corresponds to the \lq cutting\rq\ mode of deformation, as opposed to the \lq radial compression\rq\ mode of deformation, more typical with blunt wedges. This free surface formation is accompanied by a pileup of material along the sides of the punch. 

Likewise, during deformation and flow of the granular medium, the free surface evolves, with a similar pileup around the sides of the punch (see the sequence in Fig.~\ref{fig:strainAccumulation}). In order to study the evolution of the free surface, particles in a thin layer (of width $1.5d$) adjoining the $x$-axis ($y=0$) are identified at $t=0$ and followed during flow. The initial and final positions (punch penetration $Y = 0$ mm and $Y = 14.32$ mm respectively) of these particles are shown as scatter plots in Fig.~\ref{fig:freeSurface}. The particles initially form a horizontal layer, shown as black/ green triangles, in the figure. The final particle locations, shown by red/ blue circles in the figure, follow the shape of the punch, while forming an upward arch beyond $x/D = \pm 1$. The variation in thickness of this final layer reveals gaps in the surface adjoining either side of the punch. These gaps do not exist in the final material, as they are filled by particles from the bulk. Therefore, similar to metals, an equivalent cutting mode of deformation also occurs in granular materials. 

The height of the material pileup on the sides of the punch is measured from the hypothetical free surface position, shown by a dotted line in Fig.~\ref{fig:freeSurface}. This is where the free surface would be if the material did not encounter the punch. \notethis{At $x/D = \pm 4$, the two levels nearly coincide, confirming negligible flow against the side walls.}

\subsubsection{Column buckling and flow}

\begin{figure}
    \centering
    \includegraphics[scale=0.35]{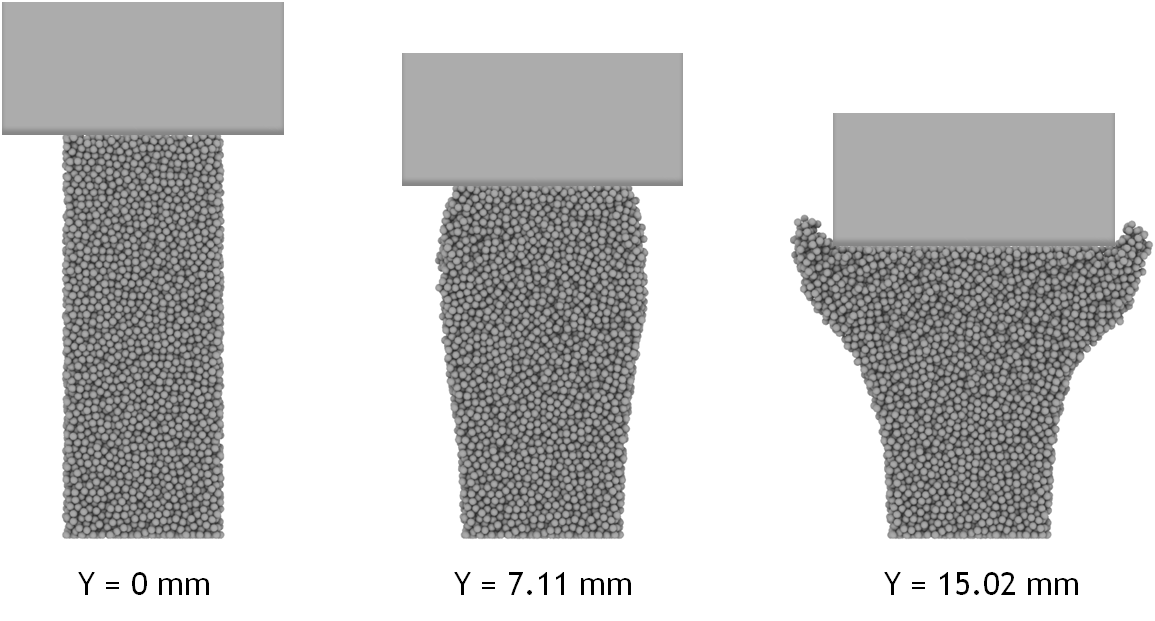}
    \caption{\label{fig:buckling} Change in shape of vertical column of particles underneath the punch. The surrounding material is not shown. The column appears to \lq buckle\rq\ before flowing past the sides of the punch.}
\end{figure}

The material immediately underneath the punch is compressed, being driven against its bottom face. To study this confined compression, a vertical column of particles under the punch is isolated. These particles are identified at $Y=0$ and their subsequent positions recorded, a rendering of which is shown in Fig.~\ref{fig:buckling}. The particles surrounding this column are not shown in the figure. Due to compression by the punch, the initially vertical (at $Y = 0$ mm) column develops a bulge (at $Y= 7.11$ mm in the figure) along its sides. With further penetration, the extent of the bulge increases, until material at the edges finally begins to flow along the sides of the punch ($Y=15.02$ mm). Beyond this stage, material particles along the punch sides continue to flow upward. These particles hence filled the gaps that appeared in the \lq top surface\rq\ of the material (see Sec.~\ref{subsubsec:freeSurface}). Thus, particles that constitute the piled up material originate both from the bulk away from the punch as well as from immediately below the punch.

The width of the initial ($Y=0$) vertical column in Fig.~\ref{fig:buckling} was chosen to coincide with the final width of the stagnation zone underneath the punch. However, it was observed that this choice was not crucial to the behavior shown in the figure --- changing the initial column width only altered the $Y$ at which flow past the sides of the punch occurred.


\section{Discussion}
\label{sec:discussion}

Our experiments on a model granular medium have shown a number of unique flow features during deformation and plastic flow. Quantitative details of these features were extracted from high resolution \emph{in situ} observations, coupled with numerical simulations. Characteristics that were not amenable to analysis by 2D imaging were extracted from the 3D NSCD simulations. 

When viewed in the rest frame of the punch, plane--strain indentation resembles flow of a dense granular material past a static obstacle. However, an important difference pertains to the role of gravity --- conventional studies of gravity--driven flow past obstacles \cite{AmaroucheneETAL_PhysRevLett_2001,TuzunNedderman_ChemEnggSci_1985,ChehataETAL_PhysFluids_2003} have involved free fall of the constituent particles. This reduces the average life of inter--particle contacts. In contrast, the flow in punch indentation occurs with multiple persistent contacts. This is because the material flow direction is against gravity. 

The primary physical constraints in our work concerned the use of monodisperse, rigid spheres, in order to eliminate the effects of particle size variation and deformation. This motivated the choice for a granular medium comprised of steel balls. The reported features were found to be qualitatively unchanged upon removing these two constraints, as seen from Movie M3 (supplemental material \cite{SuppMat}), which shows punch indentation of a medium composed of mustard seeds. These are crushable polydisperse grains, yet the observed flow pattern remains qualitatively similar to that of a medium made of rigid monodisperse particles. Several features observed in our model system have also been reported in other granular media composed of dry sand \cite{MurthyETAL_PhysRevE_2012} and elastic disks \cite{ZhangETAL_GranularMatter_2010}. The physical properties of the individual particles used in these studies (size, shape, Young's modulus and friction coefficient) are very different from the steel balls used in our experiments. These observations, taken in total, suggest universality of the observed flow patterns in granular media. 

The formation of a stagnation zone in flow past an obstacle has been reported for smaller grains \cite{MurthyETAL_PhysRevE_2012,TuzunNedderman_ChemEnggSci_1985,AmaroucheneETAL_PhysRevLett_2001}. In these cases they appear to occur only because of the existence of long-lasting interparticle contacts in a narrow region surrounding the obstacle. In our configuration, interparticle contacts are largely maintained throughout the flow, because the particles are not in free fall. The occurence of the stagnation zone and its subsequent evolution (\emph{cf.} Fig.~\ref{fig:flowFeatures}) are determined purely by interparticle rigid-body interactions and Coulomb friction at the contacts. The coefficient of friction determines the final extent of this stagnation zone, as confirmed by the close agreement between experimental results and NSCD simulations (Fig.~\ref{fig:angleTimePlot}). The occurence of a stagnation zone has implications for powder processing of metals and ceramics. In these applications, homogeneous densities are desired. In contrast, in pharmaceutical powder compaction for making tablets, a gradation in density is sought. The stagnation zone inhibits consolidated particle mixing during compression and is undesirable both for uniform and graded densities. Thus, methods to reduce its size must be employed. Our experiments suggest that this may be achieved by reducing the coefficient of friction between individual grains (see Fig.~\ref{fig:angleTimePlot}). This will reduce the base angle $\theta$, and hence the extent of the stagnation zone, enabling free mixing.

\notethis{Classical Mohr--Coulomb plasticity solutions for quasi--static deformation also predict a stagnation zone \cite{Nedderman_StaticsKinematicsGranular} beneath the punch, just prior to the onset of slip. The base angle of this zone is a function of the internal friction angle $\psi$, and is $90^\circ$ for materials with $\psi = 0$. While the present experiments and simulations have dealt with particle--level properties (such as the friction coefficient $\mu$), there is as yet no certain way to relate these to continuum--scale material constants like $\psi$ \cite{AdamsBriscoe_TribologyParticulateTech}}. 

Vortex--like circulation centers developed during the course of the flow (Fig.~\ref{fig:particleVelocityVorticity}) have also been reported in gravity--driven granular flow past obstacles  \cite{ChehataETAL_PhysFluids_2003} and as \lq eddies\rq\ in two--dimensional sheared granular materials \cite{RadjaiRoux_PhysRevLett_2002,MillerETAL_PhysRevLett_2013}. This is again a consequence of simple rigid body dynamics and friction, as seen from the NSCD simulations (\emph{cf.} Figs.~\ref{fig:velFieldExp} and \ref{fig:velFieldSim}). Interestingly, as shown in Fig.~\ref{fig:rotScatter}, vorticity in the material decays quickly with distance from the punch, reminiscent of a boundary layer in fluid flow. Such boundary layers have been reported in simulations of 2D simple shear \cite{ShojaaeeETAL_PhysRevE_2012} suggesting a correlation with large local dissipation due to inter--particle friction. These vortices may be reduced by the introduction of rounded edges on the punch, an idea that we hope to explore in the future. This could lead to the development of \lq laminar\rq\ granular flows with reduced energy dissipation, and better control of powder densities for processing applications. 

Shear band formation is common to many granular materials containing different types of grains \cite{SchallvanHecke_AnnRevFluMech_2009}. In our system, since the particles themselves were perfectly rigid, plastic strain in the material resulted from large relative motion of neighboring particles. Regions of high velocity gradients also correspond to regions of high shear strain rate --- see Fig.~\ref{fig:strainRateVariation}. As a consequence, if the kinematic configuration results in large velocity gradients (such as those marked by the arrows in Fig.~\ref{fig:strainRate}), the shear strain will tend to get localized in bands. Inside the band themselves, an increase in volume (dilatancy) can be expected and this has been reproduced in the NSCD simulations (Fig.~\ref{fig:shearBandVelocities}). However, in the present experiments, these bands developed as a consequence of competition between compaction due to the punch and material dilatancy \cite{KablaSenden_PhysRevLett_2009} --- while compaction of material immediately below the punch increases the local density, the particles themselves prefer to move apart and create free space. 

Shear banding is a common cause of failure for Bulk Metallic Glasses (BMGs). The structure of many BMGs has often been approximated by hard sphere media \cite{Spaepen_ActaMet_1977,Argon_ActaMet_1979,ShimizuETAL_MatTrans_2007}, and the occurence of shear bands has been attributed to excess free volume concentration or the collective formation of STZs \cite{FalkLanger_PhysRevE_1998}. This idea is consistent with the present observations, since existence of local pockets of free volume can result in particle jumps, thereby inducing a large velocity gradient. 

The visual similarity between the vertical column underneath the punch (Fig.~\ref{fig:buckling}) and a buckled elastic column shows how additional strain is actually accomodated. It is interesting to note that force chains in the medium also show a tendency to buckle \cite{TordesillasETAL_GranularMatter_2011} and most likely have a role to play in this process. One important distinction, however, is the reversibility in elastic column  buckling --- the deformation in Fig.~\ref{fig:buckling} is completely irreversible. While further flow in the medium occurs after the bulge formation, this change cannot be reversed, even in the absence of gravity. 

Granular flows, both in the present (mesoscopic) case and fine-grained media (e.g., sand) \cite{MurthyETAL_PhysRevE_2012}, show remarkable similarity with punch indentation of rigid-plastic metals. In metal indentation, a central stagnation zone is developed right underneath the indenter, along with shear bands bounding it --– exactly as in the present experiments. These have been predicted in metals by slip line field analyses --– the Prandtl field \cite{Hill_MathematicalTheoryPlasticity} –-- and confirmed in experiments \cite{MurthyETAL_ProcRoySocA_2014}. Likewise, this flow is reproduced in exact-scale NSCD simulations of the granular medium. As in metals, the shear bands represent regions of high local shear strain and significant velocity jumps (Fig.~\ref{fig:strainAccumulation}). The NSCD simulations also show that the formation of new surfaces when the punch penetrates the granular medium (Fig.~\ref{fig:freeSurface}) is very similar to that in metal indentation \cite{SamuelsMulhearn_JMechPhysSolids_1957}.

Given this strong similarity between indentation flows in metals and granular media, it is likely that the vortex--like structures observed in the present study also occur with metal indentation. While this is yet to be verified, perhaps, due to lack of adequate observations along indenter walls in metals, it is nevertheless promising that such circulation zones have been captured in sliding metal surfaces \cite{SundaramETAL_PhysRevLett_2012}.


\section{Conclusions}
\label{sec:conclusions}


The \emph{in situ} imaging and 2D image analysis of punch indentation of a mesoscopic, dense granular medium has allowed flow features to be studied at high resolution. Detailed information about particle positions and velocities has been obtained. The use of an analysis method for the disordered grain network has enabled the calculation of strain fields, velocity gradients and rotations. Based on this data, the formation of collective patterns in the flow has been captured. Despite its coarse discrete nature, the model granular material shows a number of flow features that bear striking resemblance to those seen in metals and fine-grained granular materials. A triangle-shaped stagnation zone is established underneath the punch, as in indentation of metals and sand. Vortices and shear bands are observed, and their characteristics have been quantified using experimentally measured deformation rates. These features are quantitatively reproduced using full-scale, 1:1 Non-Smooth Contact Dynamics (NSCD) simulation, showing also that they originately purely from local rigid body dynamics and Coulomb friction. The coupled simulation has provided clues to the kinematic origin of shear bands, by direct visualization of the dilatancy in the plane of the bands. New flow features, such as free surface evolution and buckling of a vertical column of particles underneath the punch, not easily accessible in experiments, are revealed by the NSCD. The observations in total suggest that, despite the large size of the particles, their collective dynamics very much resemble a continuum. This is perhaps why continuum theories have been quite successful in analyzing quasi-static deformation problems pertaining to granular media.

\section*{Acknowledgements}
This work was supported in part by US Army Research Office Grant W911NF-12-1-0012 and NSF grant CMMI 1234961 (Purdue); Department of Science and Technology (DST, India) grant no. SR-CE-0057-2010 (to T.G.M., Indian Institute of Science); and a Bilsland Dissertation Fellowship (to K. V., Purdue). Use of codes made available by D. Blair (Georgetown U), E. Dufresne (Yale) and A. Donev (NYU) is gratefully acknowledged. 


\bibliography{manuscript}
\bibliographystyle{unsrt}
\end{document}